\documentclass[floatfix,twocolumn,amsmath,amssymb,latin]{jpsj2}
\title{Nesting Induced Peierls-type Instability for Compressed Li-$cI16$}

\author{\textsc{Alvaro Rodriguez-Prieto}$^{1,2}$, \textsc{Viatcheslav M. Silkin}$^{2}$ and \textsc{Aitor Bergara}$^{1,2,3}$\thanks{E-mail address: a.bergara@ehu.es}}

\inst{$^1$Materia Kondentsatuaren Fisika Saila, Zientzi Fakultatea, Euskal Herriko Unibertsitatea,
644 Posta kutxatila, E-48080 Bilbo, Basque Country, Spain\\
$^2$Donostia International Physics Center (DIPC), Manuel de Lardizabal Pasealekua,
E-20018 Donostia, Basque Country, Spain\\
$^3$Centro Mixto CSIC-UPV/EHU, 1072 Posta kutxatila,
E-20080 Donostia, Basque Country, Spain\\}

\abst{Alkalies are considered to be simple metals at ambient conditions. However, recently reported theoretical and experimental results have shown an unexpected and intriguing correlation between complex structures and an enhanced superconducting transition temperature in lithium under pressure. In this article we analyze the pressure induced Fermi surface deformation in $bcc$ lithium, and its relation to the observed $cI16$ structure. According to our calculations, 
the Fermi surface becomes increasingly anisotropic with pressure and develops an extended nesting along the $bcc$ [121] direction. This nesting induces a phonon instability of both transverse modes at $N$, so that a Peierls-type mechanism 
is proposed to explain the stability of Li-$cI16$.}

\kword{Lithium, High Pressure, Fermi Surface, Nesting, Structural Transition.}

\begin{document}

\maketitle

Light alkalies have long been considered as simple metals \cite{wigner} due to their monovalency, high conductivity and crystallization in high symmetric structures. However, new theoretical and experimental results have shown that under pressure light alkalies depart radically from this simple behavior, as phase transitions to complex and low-coordinated structures emerge. \cite{NA,hanfland,CN,prb} According to a recent x-ray analysis \cite{hanfland} lithium undergoes a $bcc$ to $fcc$ transition at 7.5 GPa, followed by a $fcc$ to $hR1$ at 39 GPa and a $hR1$ to $cI16$ at around 40 GPa. On the other hand, despite experiments looking for superconductivity in lithium at equilibrium have failed \cite{liambient}, it has been observed to superconduct at 20 K when pressure rises to 40 GPa \cite{deemyad}. This surprising and intriguing correlation between complex structures and superconductivity have rosed the interest to characterize physical properties of compressed lithium. 

In this article we present the evolution under pressure of the Fermi surface (FS) topology in $bcc$ lithium, as a key ingredient explaining observed phonon instabilities, which determines a nesting induced Peierls-type mechanism as the origin of the $cI16$ structure in lithium. Electronic properties in lithium under pressure are analyzed using a plane-wave implementation of the Density Functional Theory (DFT) within the Local Density Approximation (LDA)\cite{vasp}, whereas phonon frequencies have been computed with Density Functional Perturbation Theory (DFPT).\cite{pwscf}

\begin{figure}[h]
\centering
\vspace*{0.5cm}
\includegraphics[scale=0.35]{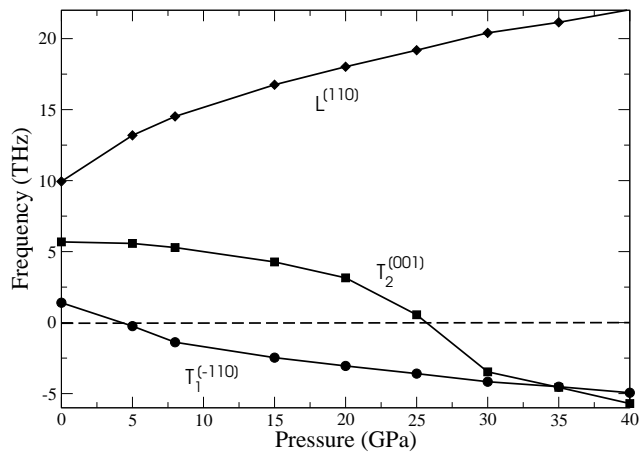}
\caption{Phonon frequencies of $bcc$ lithium at the $N$ point of the BZ boundary as a function of pressure. $L$ and $T$ labels longitudinal and transverse modes. Although the $T_1$ mode (polarized along [1$\bar{1}$0]) becomes unstable at around 5 GPa (corresponding to the $bcc$ to $fcc$ transition), the $T_2$ branch (polarized along [001]) is stable up to around 25 GPa. On the other hand, the longitudinal mode increases constantly with pressure.}
\label{fig:phonons}
\end{figure}

Li-$cI16$ structure can be easily characterized starting from a parent $bcc$ with double unit cell and considering atomic displacements along the cubic diagonal which transform as a transverse zone boundary instability at the $N$ point of the
$bcc$ Brillouin zone (BZ).\cite{katzke} Fig. \ref{fig:phonons} displays $bcc$ phonon frequencies evaluated at $N$ as a function of pressure. Although the longitudinal mode increases under pressure, at around 5 GPa, close to the $bcc$ to $fcc$ transition, the $T_1$ transverse mode (polarized along [$\bar{1}$10]) becomes unstable. In addition, when pressure rises to around 25 GPa, $T_{2}$ (polarized along [001]) also become unstable at $N$. 
In fact, as mentioned above, instability of both transverse modes is required in order to get the $cI16$ structure from a parent $bcc$, because transverse atomic displacements along the diagonal of the cube ([$\bar{1}$11]) can only be obtained by a linear combination of $T_1$ and $T_2$.

On the following, we present the evolution of the FS topology of $bcc$ lithium under pressure, as a key ingredient determining the modification of its physical properties. The nearly spherical shape it presents at equilibrium, confirming the nearly free electron character at normal conditions, significantly deforms under compression (Fig. 2). Such strong distortion of the FS under pressure can only be explained by an increasing $s$ to $p$ orbital mixing induced by a growing pressure-driven non-local effects of the ionic pseudopotential.\cite{aitor} At 5 GPa the FS  starts touching the $N$ point in the BZ boundary, which is correlated to the above mentioned instability of $T_1$ at $N$ and the $bcc$ to $fcc$ transition, explained by a Hume-Rothery mechanism.\cite{PRL} More importantly, with increasing pressure the FS develops extended nestings, so that both $T_{1}$ and $T_{2}$ become unstable at $N$ when P$>$25 GPa.

\begin{figure}[h]
\centering
\vspace*{0.5cm}
\includegraphics[scale=0.5]{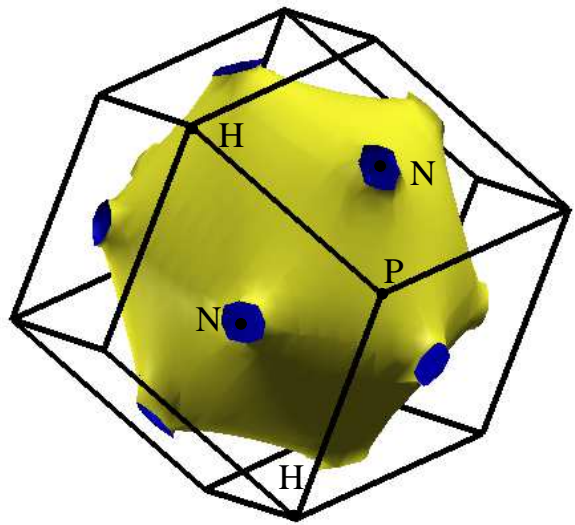}
\caption{Fermi surface of  \textit{bcc} lithium at P=40 GPa. The  spheroidal Fermi surface at equilibrium becomes increasingly distorted under pressure and at around 5 GPa, close to the experimental \textit{bcc} to \textit{fcc} transition, contacts the Brillouin at the \textit{bcc} $N$ point. With pressure the Fermi surface of  \textit{bcc} lithium shows increasing necks along $\Gamma N$ and, more interestingly, develops an extended nesting along [121], which becomes the origin of several modifications on its physical properties. }
\label{fig:FS}
\end{figure}

\begin{figure}
\centering
\includegraphics[scale=0.525]{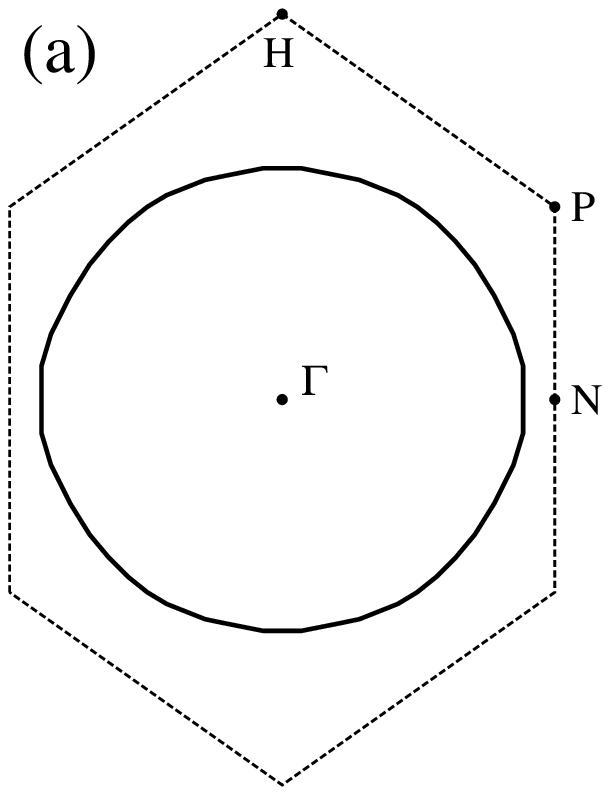}
\includegraphics[scale=0.525]{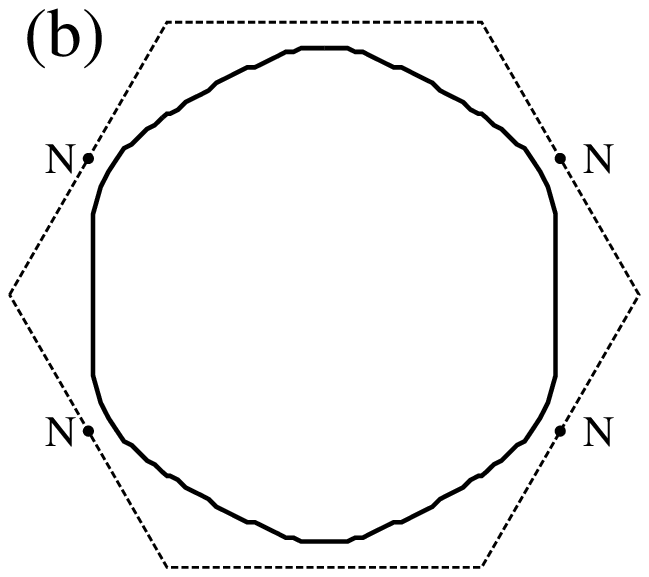}\\
\includegraphics[scale=0.525]{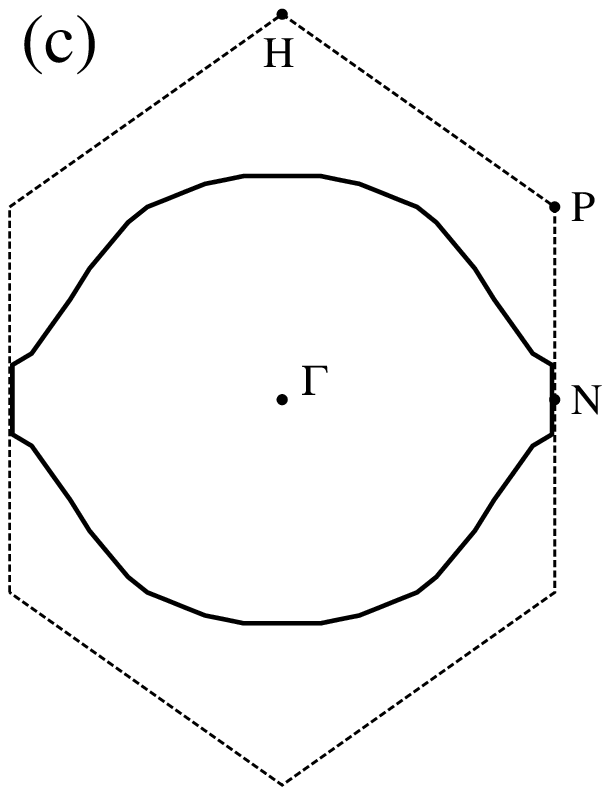}
\includegraphics[scale=0.525]{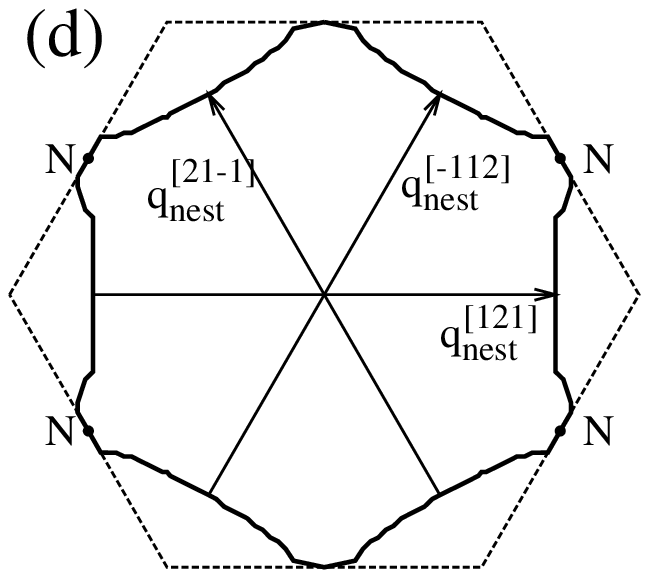}
\caption{Fermi surface cross sections (solid lines) of  \textit{bcc} lithium along (110) and (1$\bar{1}$1) planes at equilibrium [(a) and (b)] and P=40 GPa [(c) and (d)], respectively. BZ boundary is represented by dotted lines. At ambient pressure the cross section along (110) shows an almost perfect circle (a) but along the (1$\bar{1}$1) plane (b), which includes six points $N$, even at P=0 becomes slightly distorted. The distortion of the Fermi surface increases under pressure and at P=5 GPa contacts the BZ boundary at $N$. More interestingly, at 40 GPa the cross section in the (1$\bar{1}$1) plane (d) presents a clear nesting along [121] and symmetrically equivalent directions, which becomes the origin of the observed structural transition to the $cI16$.}
\label{fig:FSC}
\end{figure}

In Fig. \ref{fig:FSC} we present FS cross sections in the (110) and (1$\bar{1}$1) planes at equilibrium [(a) and (b)] and P=40 GPa [(c) and (d)], respectively. Although at equilibrium the cross section in the (110) is almost a perfect circle, it becomes slightly distorted in the (1$\bar{1}$1) plane and, following this trend,  at 40 GPa shows an extended nesting along the [121] (${\bf q}_{\rm{nest}}^{[121]}=\pi/a(1,2,1)=2k_{\rm F}$) and two other symmetrically equivalent directions. Interestingly, this nesting originates a kink in the electronic susceptibility and is connected to the observed phonon instability at $N$. However, $\Gamma N \neq {\bf q}_{\rm{nest}}^{[121]}$. In order to compare both momenta we have to take into account that ${\bf q}_{\rm{nest}}^{[121]}$ extends out from the first BZ, which after folding back with ${\bf G}=2\pi/a(-2,0,0)$ results to be the $N$ point at the BZ boundary: ${\bf q}_{\rm{nest}}^{[121]} + {\bf G}=\pi/a (1,0,-1)=\Gamma N$. Additionally, the calculated nesting vector lies exactly above the observed [211] peak of the experimental x-ray diffraction pattern corresponding to the $cI16$ phase,\cite{hanfland} which becomes relevant determining the nesting induced Peierls-type mechanism to stabilize this structure. 

%According to these experiments, the $cI16$ phase is already present at 39.8 GPa with a corresponding unit cell parameter of a$_{cI16}$= 5.2716 \AA, which doubles the structural parameter we have calculated for the bcc structure at 40 GPa, a$_{bcc}$= 2.62 \AA. 

In conclusion, we have performed \textit{ab initio} calculations of the Fermi surface deformation with increasing pressure in $bcc$ lithium. As a result of the increasing non-local character of the pseudopotential, the Fermi surface of $bcc$ lithium becomes highly anisotropic under pressure and presents an extended nesting along the [121] direction, which explains the observed phonon instabilities of both transverse modes at $N$ and determines a nesting induced Perierls-type mechanism as the origin of Li-$cI16$. It is noteworthy that we also expect interesting anomalies in the optical and dynamic screening electronic properties to arise in lithium under pressure, as dropping bands below the Fermi energy will enhance electronic interband transitions. Finally, complex structural transitions under pressure are not unique in lithium, but have been observed in heavier alkalies and other elements as well, where very similar  Peierls-type instabilities might be also expected.

The authors acknowledge partial support by the University of the 
Basque Country, the Basque Herkuntza, Unibersitate eta Ikerketa saila 
and Spanish MCyT (Grant No.FIS 2004-06490-C03-01), and the European 
Community 6th Network of Excellence NANOQUANTA (NMP4-CT-2004-500198).


\begin{thebibliography}{99} %% The number "99" means that this list has more than nine items.

\bibitem{wigner}E. Wigner and F. Seitz, \textbf{Phys. Rev.} 43 (1933) 804.

\bibitem{NA}J.B. Neaton and N. W. Ashcroft, \textbf{Nature} 400 (1999) 141.

\bibitem{hanfland}M. Hanfland, K. Syassen, N.E. Christensen and D.L. Novikov, \textbf{Nature} 408 (2000) 174.

\bibitem{CN}N.E. Christensen and D.L. Novikov, \textbf{Phys. Rev. Lett.} 86 (2003) 1881.

\bibitem{prb}A. Rodriguez-Prieto and A. Bergara, \textbf{Phys. Rev. B} 72 (2005) 125406. 

\bibitem{aitor}A. Bergara, J.B. Neaton and N.W. Ashcroft, \textbf{Phys. Rev. B} 62 (2000) 8494. 

\bibitem{katzke}H. Katzke and P. Toledano, \textbf{Phys. Rev. B} 71 (2005) 184101.

\bibitem{liambient}K.I. Juntunen and J.T. Tuoriniemi, \textbf{Phys. Rev. Lett.} 93 (2004) 157201.

\bibitem{deemyad} K. Shimizu {\it et al.}, \textbf{Nature} 419 (2002) 597; V.V. Struzhkin {\it et al.}, \textbf{Science} 298 (2002) 1213; S. Deemyad and J.S. Shilling, \textbf{Phys. Rev. Lett.} 91 (2003) 167001; N.W. Ashcroft, Nature 419 (2002) 569.

\bibitem{vasp}G. Kresse and J. Hafner, \textbf{Phys. Rev. B} 23 (1981); G. Kresse and J. Furthm\"uller, \textbf{Phys. Rev. B} 54 (1996) 11169.

\bibitem{pwscf}S. Baroni {\it et al}, http://www.pwscf.org.

\bibitem{PRL}A. Rodriguez-Prieto, V.M. Silkin, A. Bergara and P.M. Echenique, submitted to \textbf{Phys. Rev. Lett.}; A. Rodriguez-Prieto and A. Bergara, Proceedings of Joint 20th AIRAPT-43rd EHPRG 2005 Conference, cond-mat/0505619.

\end{thebibliography}
\end{document}